\documentclass[aps,prl,amsmath,showpacs,twocolumn]{revtex4}

\usepackage[latin1]{inputenc}
\usepackage[T1]{fontenc}
\usepackage[dvips]{graphicx}

\date{2008-01-24}

\newcommand{\br}[1]{\langle #1\vert}
\newcommand{\ke}[1]{\vert#1\rangle}
\newcommand{\bk}[2]{\langle #1|#2\rangle}

\newcommand{\bok}[3]{\br{#1}#2\ke{#3}}

\begin{document}
\title{Montonic convergent optimal control theory to modulate bandwidth limited laser pulses in
linear and non-linear optical processes}
\author{Caroline Gollub, Markus
Kowalewski and Regina de Vivie-Riedle}

\affiliation{Department of Chemistry,
Ludwig-Maximilians-Universit\"at M\"unchen, D-81377 M\"unchen,
Germany}

\pacs{33.80.Wz, 03.67.Lx, 02.30.Yy}

\begin{abstract}
We present a modified optimal control scheme based on the Krotov method,
which allows for strict limitations on the spectrum of the optimized
laser fields, without losing monotonic convergence of the algorithm.
The method guarantees a close link to learning loop control experiments and
is demonstrated for the challenging control of non-resonant Raman transitions,
which are used to implement a set of global quantum gates for molecular vibrational qubits.
\end{abstract}

\maketitle

With the progress of laser pulse shaping and learning loop techniques~\cite{Brixner07}  quantum control experiments (OCE) became a forefront tool for the control and deciphering of molecular quantum processes~\cite{Assion98, Rabitz00, Motzkus02, Woeste03}. Optimal control theory (OCT)~\cite{Tannor97, Rabitz98}  as the theoretical counterpart is a powerful method for the prediction of pulse structures as initial guess and guidance for OCE. With OCT, insight into the quantum pathways of these processes is directly available. The numerous applications of optimal control range from the control of chemical reactions in gas and condensed phase~\cite{Gerber03,Levis02} to the control in nanostructures~\cite{BrixnerNF07, Hoefer07} and to quantum optical problems like quantum information processing~\cite{Tesch02, Troppmann07, Glaser07, Zoller} or the preparation of cold molecules~\cite{Koch04, Tannor99}.

One fundamental difference between OCE and OCT is the spectral bandwidth of the laser field inherently present in the experiment but in principle unlimited
in the original theoretical formulation. The general comparability of experimental and theoretical results may be complicated, since the theoretical answer for the optimal pulse can always span a wide bandwidth with quantum pathways out of experimental reach. Several suggestions have been made dealing with this challenge~\cite{Gross92, Artamonov04, Werschnik05}, however, at the cost of monotonic convergence or general applicability~\cite{Thomas}.

We present a modified OCT approach based on the Krotov method~\cite{Somloi} that treats time and frequency domain equally while providing monotonic convergence. The method offers an elegant possibility to study OCEs theoretically by explicitly including as a constraint the crucial experimental feature of the spectral bandwidth.

As an ultimate test we demonstrate the new tool by the implementation of stimulated non-resonant Raman quantum gates for vibrational qubits. The idea of molecular vibrational quantum computing~\cite{Tesch02} has first been introduced for IR-active modes. Ultrashort, specially shaped laser pulses act as global quantum gate operations.  Different types of quantum gates and quantum algorithms have been demonstrated theoretically for IR transitions~\cite{Tesch02, NJP1, NJP2, Brown06, Desouter07} and STIRAP processes~\cite{DesouterSTIRAP07,Shapiro}. Experimentally, molecular quantum gates have been realized in the visible regime~\cite{Apkarian01, Leone02} and latest IR shaping experiments~\cite{Zanni1, Zanni2} open the route for the realization of vibrational qubits in the IR~\cite{Korff05}. Stimulated non-resonant Raman quantum gates will provide new flexibilities, like the choice of laser wavelengths. Their theoretical implementation can be regarded as a great challenge for the new OCT scheme since it comprises a non-resonant, two-photon, two-color process.

The multi-target optimal control functional~\cite{Tesch02} for the molecular non-resonant Raman interaction includes two laser fields $\epsilon_l(t)$ with $l=1,2$ (Eq.~\ref{OCT}). The corresponding Raman Hamiltonian is given by $V_R=-\frac{1}{2}\epsilon_1(t)\,\hat\alpha\,\epsilon_2(t)$, with the molecular polarizability $\hat{{\alpha}}$.
\begin{widetext}
\begin{equation}
\begin{split}
J[\Psi_{k}(t),\Phi_{k}(t),\epsilon_1(t),\epsilon_2(t)] &= {\bf \sum_k} \Bigg \lbrace | \langle{\Psi_{k}(T)} |{\Phi_{k}}\rangle | ^2
- \sum_{l=1}^2 \alpha_0 \int_0^T \frac{|\epsilon_l(t)-\tilde\epsilon_l(t)|^2}{s(t)} dt
- \sum_{l=1}^2 \gamma_l|F_l(\epsilon_l(t))| \\
&- \;2\; \Re \Bigg[ \langle {\Psi_{k}(T)}|\Phi_{k}\big\rangle \int_0^T
\bok{\Phi_{k}(t)}{\left[\frac{i}{\hbar} \left(\hat H_0 - \frac{1}{2}\, \epsilon_1(t)\,\hat\alpha\,\epsilon_2(t) \right)+ \frac {\partial}{\partial t}\right]} {{\Psi_{k}(t)}} dt \Bigg]  \Bigg\}
 \label{OCT}
\end{split}
\end{equation}
\end{widetext}
The objective is determined by the square of each overlap $\langle {\Psi_{k}(T)} |{\Phi_{k}}\rangle$  of the propagated initial states $\Psi_{k}(T)$ at the final time $T$ with the target states $\Phi_{k}$ of the global quantum gate operation. The change of the pulse energy is restricted with the Krotov-change parameter $\alpha_0$. $\tilde\epsilon_l(t)$ are the reference fields. A temporal shape function~\cite{Sundermann99}  denoted as $s(t)$ is inserted to achieve smooth switching on and off behavior of each laser field.  The wave function has to satisfy the time dependent Schr\"odinger equation including the time evolution of the non-resonant Raman process. With the new term $F_l(\epsilon_l(t))$ we introduce a  frequency filter operation  in its time representation, which restricts each electric field $\epsilon_l (t)$.
The corresponding Lagrange multipliers are $\gamma_l(t)$.
The filter operations can in principle be realized in the time domain by
linear digital filters, and particularly by finite impulse response (FIR)
filters~\cite{OppenheimSchaffer}
\begin{equation}
F(\epsilon(t)) = \sum_{j=0}^N c_j \epsilon(t-j \Delta t),\nonumber
\end{equation}
with the FIR filter coefficients $c_j$ and the
step size $\Delta t$  in the discrete time representation.
By variation of the functional (Eq.~\ref{OCT}) with respect to the initial states $\Psi_{k}(t)$, the target states $\Phi_{k}(t)$ and the laser fields $\epsilon_l(t)$ a set of coupled differential equations can be derived. The iterative calculation of the laser fields is performed with the Krotov method~\cite{Somloi} and the next iteration step $n+1$ for the laser field $\epsilon_1(t)$ and analogously for $\epsilon_2(t)$ can be formulated as:
\begin{gather}
\begin{split}
\label{eq:e_new2}
\epsilon^{n+1}_1(t) = ~ & \epsilon^{n}_1(t) - \frac{s(t)}{2\alpha_0} \left(
\gamma_1(t) -  \sum_k C_{1,k} \right)
\end{split}\\
\begin{split}
\text{with~}
\epsilon_1^n(t)=\tilde\epsilon_1(t)\nonumber
\end{split}\\
\begin{split}
\label{eq:e_new3}
\text{and}\\
C_{1,k} =~&\Im[ \bk{\Phi_k(t,\epsilon_1^{n},\epsilon_2^{n})}
{\Psi_k(t,\epsilon_1^{n+1},\epsilon_2^{n+1})} \\
&\times \br{\Phi_k(t,\epsilon_1^{n},\epsilon_2^{n})}
\hat \alpha \epsilon^{n+1}_2 \ke{\Psi_k(t,\epsilon_1^{n+1},\epsilon_2^{n+1})}]{\text{.}}
\end{split}
\end{gather}
The Lagrange multipliers $\gamma_l(t)$ can be interpreted as correction fields needed to suppress the undesired frequency components. In the optimal case the Lagrange multipliers $\gamma_l (t)$ are adjusted to substract exactly the undesired field components from the optimized uncorrected fields $\sum_k C_{l,k}$ (Eq.~\ref{eq:e_new3}). The spectral constraint $F_l(\epsilon_l(t))$ depends only linearly on each electric field and it is possible to realize the side conditions $|F_l(\epsilon_l(t))|=0$ using Fourier filters $f_l(\omega)$. It turned out that under practical considerations it is easier to use Fourier filters instead of FIR filters. The Lagrange side conditions can be implemented in form of band-stop filter operations using the inverse $f_l'(\omega) = 1-f_l(\omega)$ of the band-pass filters
$f_l(\omega)$
which guarantees that only the undesired spectral components pass the band-stop filter. The Lagrange multipliers $\gamma_l(t)$ cannot be determined directly. In fact, for the calculation of  $\gamma_l(t)$ the field change $\sum_k C_{l,k}$ must be predicted in the actual iteration step.  This task is performed by propagating the target states $\Phi_{k}$  and the intital wave functions $\Psi_{k}$ with the laser fields $\epsilon^{n}_l(t)$  from the previous iteration. 
The construction of the resulting fields $\gamma'_l(t)$ resembles the OCT fields of the unmodified algorithm.
\begin{equation}
\begin{split}
\gamma_1'(t) = & \sum_k \Im[ \bk{\Phi_k(t,\epsilon_1^{n},\epsilon_2^{n})}{\psi_k(t,\epsilon_1^{n}, \epsilon_2^{n})} \\
 \times \br{&\Phi_k(t,\epsilon_1^{n}, \epsilon_2^{n})}
\hat \alpha \epsilon_{2}^n \ke{\psi_k(t,\epsilon_1^{n}, \epsilon_2^{n})}]\approx \sum_k C_{1,k}\nonumber
\end{split}
\end{equation}
Filtering this output  $\gamma'_1(t)$ or analogously  $\gamma'_2(t)$ with the band-stop filter $f_l'(\omega)$ transforms them into the correction fields $\gamma_1(t)$ or $\gamma_2(t)$. The transformation is accomplished with the help of Fourier transforms ${\cal F}$.
\begin{equation}
 \label{eq:gamma}
 \gamma_l(t) = {\cal F}^{-1}[f_l'(\omega) \cdot {\cal F}(\gamma'(t))]
\end{equation}
Each of the new fields $\epsilon_l^{n+1}(t)$ can now be calculated by inserting the result from Eq.~\ref{eq:gamma} in Eq.~\ref{eq:e_new2}. The Lagrange multipliers $\gamma_l(t)$ represent the time dependent electric fields of the undesired frequency components. The correction fields are evaluated in each iteration step and are substracted  from the optimized uncorrected field (Eq.~\ref{eq:e_new2}). Finally, to maintain the validity of the side condition the optimized field has to be filtered with the band-pass $f_l(\omega)$ after each iteration.

The modified OCT-scheme provides monotonic convergence, i.e. each iteration step improves the objective. Its convergence is proved in close analogy to the procedure given in~\cite{Koch04} for the standard Krotov OCT.
The difference in the line of argumentation arises from the new constraints and enters in:
\begin{equation}
\label{eq:proof}
\begin{split}
  \int_0^T &- \frac{\alpha_0}{s(t)} \Delta\epsilon_l^2(t)
	+ \gamma_l(t)\,| F_l(\epsilon_l^{n+1}(t)) |
	- \gamma_l(t)\,| F_l(\epsilon_l^{n}(t)) | \\
        &+ \Delta \epsilon_l(t) \left[ 2 \frac{\alpha_0}{s(t)} \Delta \epsilon_l(t) + \gamma_l(t) \right] dt \geq  ~0,
\end{split}
\end{equation}
corresponding to Eq.~A15 of~\cite{Koch04}. $\Delta \epsilon_l(t) = \epsilon_l^{n+1}(t) - \epsilon_l^{n}(t)$ denotes the change of the laser fields between two iterations.

In accordance with a Lagrange side condition the output of the filter operations has to be zero. Consequently, all terms of Eq.~\ref{eq:proof} containing $F_l(\epsilon(t))$ vanish. Inserting Eq.~\ref{eq:e_new2} into Eq.~\ref{eq:proof} leads to:
\begin{equation}
\begin{split}
\int_0^T & \frac{3}{4}\frac{s(t)}{\alpha_0} \gamma_l^2(t)
+ \frac{ s(t)}{\alpha_0} \gamma_l(t) \sum_k C_k \\
&+ \frac{1}{4} \frac{s(t)}{\alpha_0} \left( \sum_k C_k \right)^2 dt\nonumber
\geq 0,
\end{split}
\end{equation}
where only one non-quadratic term appears besides two positive, quadratic ones. As a result, the integral is always greater than or equal to zero, which meets the requirements of convergency.
\begin{figure}[ht]
  \begin{center}
    \includegraphics[clip,width=0.68\columnwidth]{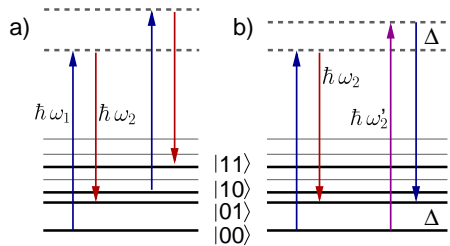}
    \caption{Stimulated non-resonant Raman quantum gates with the four two-qubit basis states. a) The exact manipulation of all quantum transitions indicated by the arrows ($|00\rangle\leftrightarrow |01\rangle$ and $|10\rangle\leftrightarrow |11\rangle$) corresponds to a global NOT gate operation. A CNOT gate is realized by pulses only  switching the state of the second (active) qubit when  the first (control) qubit is in state $|1\rangle$. b) OCT optimization of a single Raman field without frequency restrictions leads to a spectrum with an additional carrier frequency $\omega_2'$ separated by $2\Delta$ with respect to $\omega_2$.}
  \label{proc}
  \end{center}
\end{figure}
We demonstrate the strength of the proposed OCT-algorithm with the implementation of highly efficient stimulated non-resonant Raman quantum gates. A schematic sketch of the vibrational ladder and the two quantum gate operations, NOT and controlled-NOT (CNOT), is depicted in Fig.~\ref{proc}\,(a). Standard OCT-schemes fail for the simultaneous optimization of two non-resonant laser fields, since the virtual states are not determined within the formalism and the carrier frequencies of the laser pulses are independent of the eigenvalues of the system. Moreover, the electric fields $\epsilon_1(t)$ and $\epsilon_2(t)$ are not distinguishable. When one laser field of the Raman process is kept constant, the frequency unrestricted optimization of the other field yields a second band $\omega_2'$ in its spectrum (Fig.~\ref{proc}\,(b)). The additional band is shifted to higher frequencies by two times the energy difference $\Delta$ of the addressed qubit basis states. Such a broad spectrum requires extremely short laser pulses.  The new algorithm provides the opportunity to optimize both laser fields within a selected and limited frequency range and simple structured, stimulated non-resonant Raman quantum gates with high efficiencies could be predicted for the first time.

As a model system we selected two strongly Raman active C-H stretching vibrations of {\it n}-butylamine. The potential energy surface as well as the polarizability tensor components were calculated with density functional theory (b3lyp/6-31++G**)~\cite{G03} along both modes. The eigenfunctions and eigenvalues were explicitly evaluated by a relaxation method~\cite{Sundermann99}. The quantum dynamics were carried out with a Chebychev propagation scheme~\cite{cheby}.
Both selected modes with the fundamental frequencies $\nu_1=$~2990\,cm\,$^{-1}$ and $\nu_2=$\,3030\,cm\,$^{-1}$ provide high but balanced anharmonicities (intramode $\Delta_1$\,=\,74\,cm\,$^{-1}$, $\Delta_2$\,=\,103\,cm\,$^{-1}$ and intermode $\Delta_{12}$\,=\,22\,cm\,$^{-1}$), which are favorable molecular properties for vibrational quantum computing~\cite{NJP1}.
For the definition of the two-qubit basis ($|00\rangle$, $|01\rangle$, $|10\rangle$, $|11\rangle$)  as sketched in  Fig.~\ref{proc}\,(a) we encode the vibrational ground state of each selected normal mode as the logic value 0 and the first excited state as the logic value 1.
\begin{figure}[t]
  \begin{center}
    \includegraphics[clip,width=0.68\columnwidth]{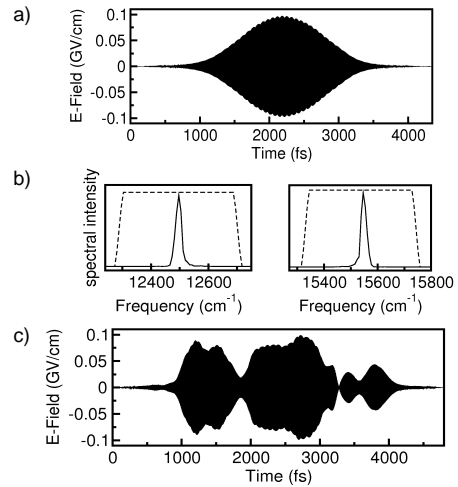}
    \caption{The Raman fields $\epsilon_1(t)$ are depicted, $\epsilon_2(t)$ have the same envelope functions, but different carrier frequencies. 
a) Laser field for the global CNOT.
b) Spectra of both CNOT fields with their respective band-pass functions (dashed lines).
c) Laser field for the global NOT.}
    \label{gate}
  \end{center}
\end{figure}
The OCT-calculations were performed in the eigenstate representation, using the 50 lowest eigenstates. For the description of the laser-molecule-interaction, we selected the $x^2$-tensor component surface and evaluated the corresponding matrix elements. A universal set of quantum gates is implemented for this two-qubit system by stimulated non-resonant Raman processes. The CNOT and NOT gate with efficiencies above 99~$\%$ are exemplarily presented. $\epsilon_1(t)$ and $\epsilon_2(t)$ of the global CNOT gate can be realized by simple gaussian-shaped laser fields (compare Fig.~\ref{gate}\,(a) for $\epsilon_1(t)$). Their related spectra are depicted together with their band-pass filter functions~$f_l(\omega)$ in Fig.~\ref{gate}\,(b). The carrier frequencies were chosen to be in the near IR-regime with 800\,nm (12500\,cm$^{-1}$) and 643\,nm (15541\,cm$^{-1}$).
\begin{figure}[b]
  \begin{center}
    \includegraphics[clip,width=0.68\columnwidth]{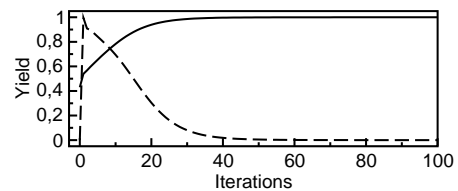}
    \caption{The solid line indicates the convergence of the modified OCT algorithm after the total time $T$ (determined by the yield $\sum_{k=1}^4 \frac{1}{4} \langle \Psi_{ik} (T) | \Psi_{fk} (T) \rangle $) during the optimization. The amount of the suppressed frequency components during the optimization is shown by the dashed line (normalized to unity).}
    \label{suppYield}
  \end{center}
\end{figure}
The global NOT operation (Fig.~\ref{gate}\,(c)) is more complex in structure, because two vibrational transitions  (compare Fig.~\ref{proc}\,(a)) have to be driven simultaneously and the fundamental qubit transition is close to the corresponding passive one. The $x^2$-tensor component of the polarizability drives the vibrational transitions for both qubit modes. Thus, it can be expected that polarized fields shaped to discriminate the qubit modes, might further simplify the laser field structures.

The monotonic convergence for the CNOT gate optimization with a Krotov-change parameter $\alpha_0$=10 can be traced from Fig.~\ref{suppYield}, solid line. The dashed line shows the evolution (normalized to unity) of the undesired spectral components during the optimization.
Since the guess fields were chosen as simple, bandwidth tailored gaussian shaped laser fields, no frequency components have to be suppressed. In the first iteration step the amount of undesired spectral components jumps to a maximum of less than 1\,$\%$ of the pulse energy, but converges to zero while reaching the optimization aim.

In summary, we present a Krotov OCT approach, which treats time and frequency domain equally, thus, unifying global optimal control with spectral constraints. 
The new tool optimizes laser fields under realistic experimental spectral conditions. Optimal laser fields and control pathways in the experimentally accessible search space will be predicted. Additionally, an arbitrary pattern can be imprinted on the selected frequency range to suppress or enhance distinct quantum pathways. Thus, a strong and direct link to OCE is provided.
The method has been successfully demonstrated for a  multi-photon process and can easily be transferred to linear processes.
A new realization strategy for molecular vibrational quantum computing has been presented by the implementation of simple structured, non-resonant stimulated Raman quantum gates of high efficiencies. Thereby, vibrational quantum computing in the ground state is transferred for the first time to the 800\,nm regime, which is well established and accessible for pulse shaping techniques.

\begin{acknowledgments}
The Deutsche Forschungsgemeinschaft through the excellence cluster Munich Centre for Advanced Photonics and through the Normalverfahren is acknowledged.
\end{acknowledgments}


\begin{thebibliography}{99}

\bibitem{Brixner07}
P. Nuernberger, G. Vogt, T. Brixner, G. Gerber, Phys. Chem. Chem. Phys,  {\bf 9}, 2470 (2007).

\bibitem{Assion98}
A. Assion, T. Baumert, M. Bergt, T. Brixner, B. Kiefer, V. Seyfried, M. Strehle, G. Gerber, Science {\bf 282}, 919 (1998).

\bibitem{Rabitz00}
H. Rabitz, R. de Vivie-Riedle, M. Motzkus, K. Kompa, Science {\bf 288}, 824 (2000).

\bibitem{Motzkus02}
J.L. Herek, W. Wohlleben, R.J. Cogdell, D. Zeidler, M. Motzkus, Nature {\bf 417}, 533 (2002).

\bibitem{Woeste03}
C. Daniel, J. Full, L. Gonz\'{a}lez, C. Lupulescu, J. Manz, A. Merli, S. Vajda, L. W\"oste, Science {\bf 299}, 536 (2003).


\bibitem{Tannor97}
A. Bartana, R. Kosloff, D.J. Tannor, J. Chem. Phys. {\bf 106}, 1435 (1997).
 
\bibitem{Rabitz98}
W. Zhu, J. Botina, H. Rabitz, J. Chem. Phys. {\bf 108}, 1953 (1998).

\bibitem{Gerber03}
T. Brixner, G. Gerber, Chem. Phys. Chem. {\bf 4}, 418 (2003).

\bibitem{Levis02}
R.J. Levis, H.A. Rabitz, J. Phys. Chem. A {\bf 106}, 6427 (2002).

\bibitem{BrixnerNF07}
 M. Aeschlimann, M. Bauer, D. Bayer, T. Brixner, F.J. Garc\'{\i}a de Abajo, W. Pfeiffer, M. Rohmer, C. Spindler, F. Steeb, Nature {\bf 446}, 301 (2007).

\bibitem{Hoefer07}
J. G\"udde, M. Rohleder, T. Meier, S.W. Koch, U. H\"ofer, Science {\bf 318}, 1287 (2007).


\bibitem{Tesch02}
C.M. Tesch, R. de Vivie-Riedle, Phys. Rev. Lett. {\bf 89}, 157901 (2002).

\bibitem{Troppmann07}
R. de Vivie-Riedle, U. Troppmann, Chem. Rev. {\bf 107}, 5082 (2007).

\bibitem{Glaser07}
A. Sp\"orl, T. Schulte-Herbr\"uggen, S.J.  Glaser, V. Bergholm, M.J. Storcz, J. Ferber, F.K. Wilhelm, Phys. Rev. A  {\bf 75}, 012302 (2007).

\bibitem{Zoller}
U. Dorner, T. Calarco, P. Zoller, A. Browaeys, P. Grangier, J. Opt. B: Quantum Semiclass. Opt.  {\bf 7}, S341 (2005).

\bibitem{Koch04}
C.P. Koch, J.P. Palao, R. Kosloff, F. Masnou-Seeuws, Phys. Rev. A {\bf 70}, 013402 (2004).

\bibitem{Tannor99}
D.J. Tannor, A. Bartana, J. Phys. Chem. A {\bf 103}, 10359 (1999).

\bibitem{Artamonov04}
M. Artamonov, T.-S. Ho, H. Rabitz, Chem. Phys.  {\bf 305}, 213 (2004).

\bibitem{Gross92}
P. Gross, D. Neuhauser, H. Rabitz, J. Chem. Phys. {\bf 96}, 2834 (1992).

\bibitem{Werschnik05}
J. Werschnik,  E.K.U. Gross, J. Opt. B: Quantum Semiclass. Opt.  {\bf 7}, S300 (2005).
 
\bibitem{Thomas}
T. Hornung, M. Motzkus, R. de Vivie-Riedle, J. Chem. Phys. {\bf 115}, 3105 (2001).

\bibitem{Somloi}
J. Soml\'{o}i, V.A. Kazakov, D.J. Tannor, Chem. Phys. {\bf 172}, 85 (1993).

\bibitem{NJP1}
C. Gollub, U. Troppmann, R. de Vivie-Riedle, New J. Phys. {\bf 8}, 48  (2006).

\bibitem{NJP2}
U. Troppmann, C. Gollub, R. de Vivie-Riedle, New J. Phys. {\bf 8}, 100  (2006).

\bibitem{Brown06}
T. Cheng, A. Brown, J. Chem. Phys. {\bf 124}, 034111 (2006).

\bibitem{Desouter07}
M. Ndong, D. Lauvergnat, X. Chapuisat, M. Desouter-Lecomte, J. Chem. Phys. {\bf 126}, 244505 (2007).

\bibitem{DesouterSTIRAP07}
D. Sugny, M. Ndong, D. Lauvergnat, Y. Justum, M. Desouter-Lecomte, J. Photochem. Photobiol. A {\bf 190}, 359 (2007).

\bibitem{Shapiro}
C. Menzel-Jones, M. Shapiro, Phys. Rev. A {\bf 75}, 052308 (2007). 


\bibitem{Apkarian01}
R. Zadoyan, D. Kohen, D.A. Lidar, V.A. Apkarian, Chem. Phys. {\bf 266}, 323 (2001).

\bibitem{Leone02}
J. Vala, Z. Amitay, B. Zhang, S.R. Leone, R. Kosloff, Phys. Rev. A {\bf 66}, 062316 (2002).

\bibitem{Zanni1}
S-H. Shim, D.B. Strasfeld, M.T. Zanni,  Optics Express  {\bf 14}, 13120 (2006).

\bibitem{Zanni2}
D.B. Strasfeld, S-H. Shim, M.T. Zanni, Phys. Rev. Lett. {\bf 99}, 038102 (2007).

\bibitem{Korff05}
B.M.R. Korff, U. Troppmann, R. de Vivie-Riedle, J. Chem. Phys. {\bf 123}, 244509 (2005).

\bibitem{Sundermann99}
K. Sundermann, R. de Vivie-Riedle, J. Chem. Phys. {\bf 110}, 1896 (1999).

\bibitem{OppenheimSchaffer}
A.V. Oppenheimer, R.W. Schaffer, J.R. Buck, \textit{Discrete Time Signal Processing}
second editon, Prentice Hall (Upper Saddle River, NJ 1999).

\bibitem{G03}
M.J. Frisch et al., GAUSSIAN03, revision D.01, Gaussian, Inc., Wallingford, CT, 2004.

\bibitem{cheby}
T.H. Ezer, R. Kosloff, J. Chem. Phys.  {\bf 81}, 3967 (1984).


\end{thebibliography}
\end{document}